# VISIBLE STARS AS APPARENT OBSERVATIONAL EVIDENCE IN FAVOR OF THE COPERNICAN PRINCIPLE IN THE EARLY 17$^{TH}$ CENTURY

Christopher M. Graney
Jefferson Community College
1000 Community College Drive
Louisville, Kentucky, USA 40272
www.jefferson.kctcs.edu/faculty/graney
christopher.graney@kctcs.edu

ABSTRACT
The Copernican Principle (which says the Earth and sun are not unique) should have observational consequences and thus be testable.  Galileo Galilei thought he could measure the true angular diameters of stars with his telescope; according to him, stars visible to the naked eye range in diameter from a fraction of a second to several seconds of arc.  He used this and the Copernican Principle assumption that stars are suns as a method of determining stellar distances.  The expected numbers of naked eye stars brighter than a given magnitude can be calculated via Galileo's methods; the results are consistent with data obtained from counting naked eye stars.  Thus the total number of stars visible to the naked eye as a function of magnitude would appear to Galileo to be data supporting the Copernican Principle.

1. INTRODUCTION

The Earth, and even the Earth's sun, is not unique, and thus is not the center of the universe. This concept is often referred to as the "Copernican Principle".[1] The Principle was promoted by Galileo and used by him in determining distances to stars. The Principle has observational consequences, and therefore should be testable. This paper will discuss how the numbers of naked-eye stars as a function of magnitude can serve as such a test. It will show how Galileo's methods can be used to calculate expected star numbers, yielding results consistent with data obtained by counting naked-eye stars, apparently confirming the Copernican Principle. It will close with a discussion of exactly what is confirmed and what is not, and some comments on the value of such an exercise for understanding Galileo and his backing of the Copernican heliocentric theory.

2. GALILEO AND THE COPERNICAN PRINCIPLE

The Copernican Principle did not spring directly from Copernicus. Copernicus's 1543 *De Revolutionibus Orbium Coelestium* hypothesizes that the Earth is one of the planets; it is not unique and not the center of the universe. However, Copernicus places a sun-centric sphere of immobile stars, a "firmament", beyond the orbit of Saturn (Figure 1).  The stars and the sun share immobility, but that is all that is explicitly stated.  The modern ideas that the stars extend into space and that interstellar distances are large arose in the decades following the publication of *De Revolutionibus* (Figure 2).  Galileo Galilei promoted the Copernican Principle in his 1632 *Dialogue Concerning the Two Chief World Systems*, asserting that the stars are "so many suns" (p. 327) distributed through space so that "some are two or three times as remote as others" (p. 382).  Thus in Galileo's view the sun is one among many suns, one among many stars, and is no more unique than the Earth.

---

1 The specific phrase "Copernican Principle" was brought into modern usage by Hermann Bondi, and later Stephen Hawking and George Ellis, to describe a universe that appears the same in all directions (Danielson 2009). The phrase would not necessarily have been recognizable to Galileo, but serves well here.

Modern discussions of his major astronomical work tend to either overlook Galileo's views and discoveries concerning the stars, or to limit themselves to his description of them in his first publication of his telescopic observations, the *Starry Messenger* (1610).[2]  In the *Starry Messenger* Galileo describes his early impression of stars as seen through the telescope:

> The fixed stars are never seen to be bounded by a circular periphery, but rather have the aspect of blazes whose rays vibrate around them and scintillate a great deal.  Viewed with the telescope they appear of a shape similar to that which they present to the naked eye, but sufficiently enlarged so that a

---

[2] As an example, the *Cambridge Companion to Galileo* contains articles on "Galileo's Copernicanism" and "Galileo's discoveries with the telescope and their evidence for the Copernican theory."  These discuss in detail Galileo's observations of the moon, Jupiter's satellites, Venus' phases, and sunspots, but mention the stars only briefly, and only in terms of the *Starry Messenger*.

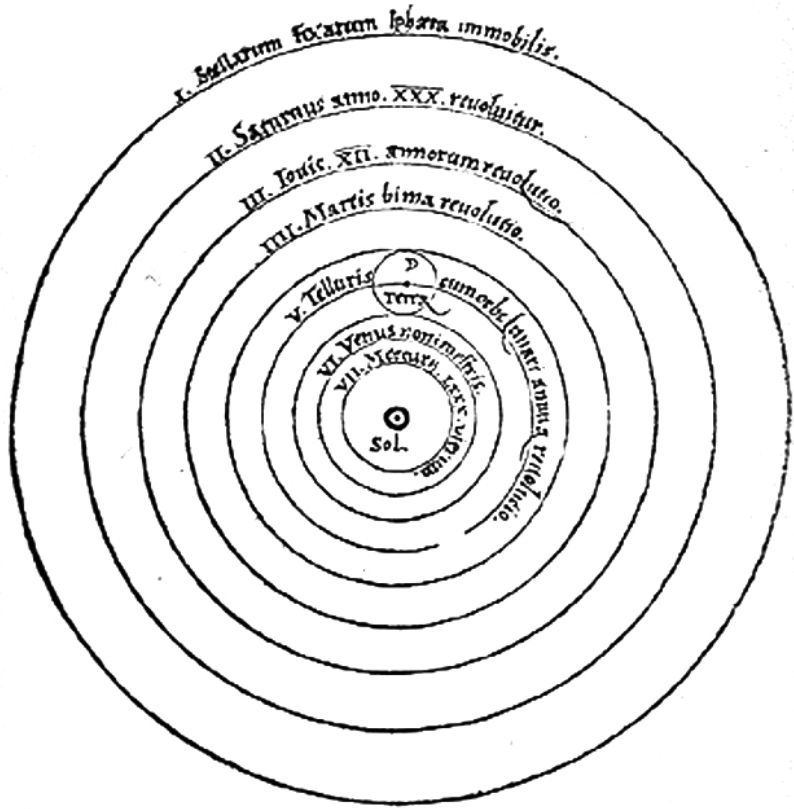

FIGURE 1: The Copernican System as portrayed in Copernicus's *De Revolutionibus Orbium Coelestium* (1543). The outer circle is the immobile sphere of the stars -- a firmament.

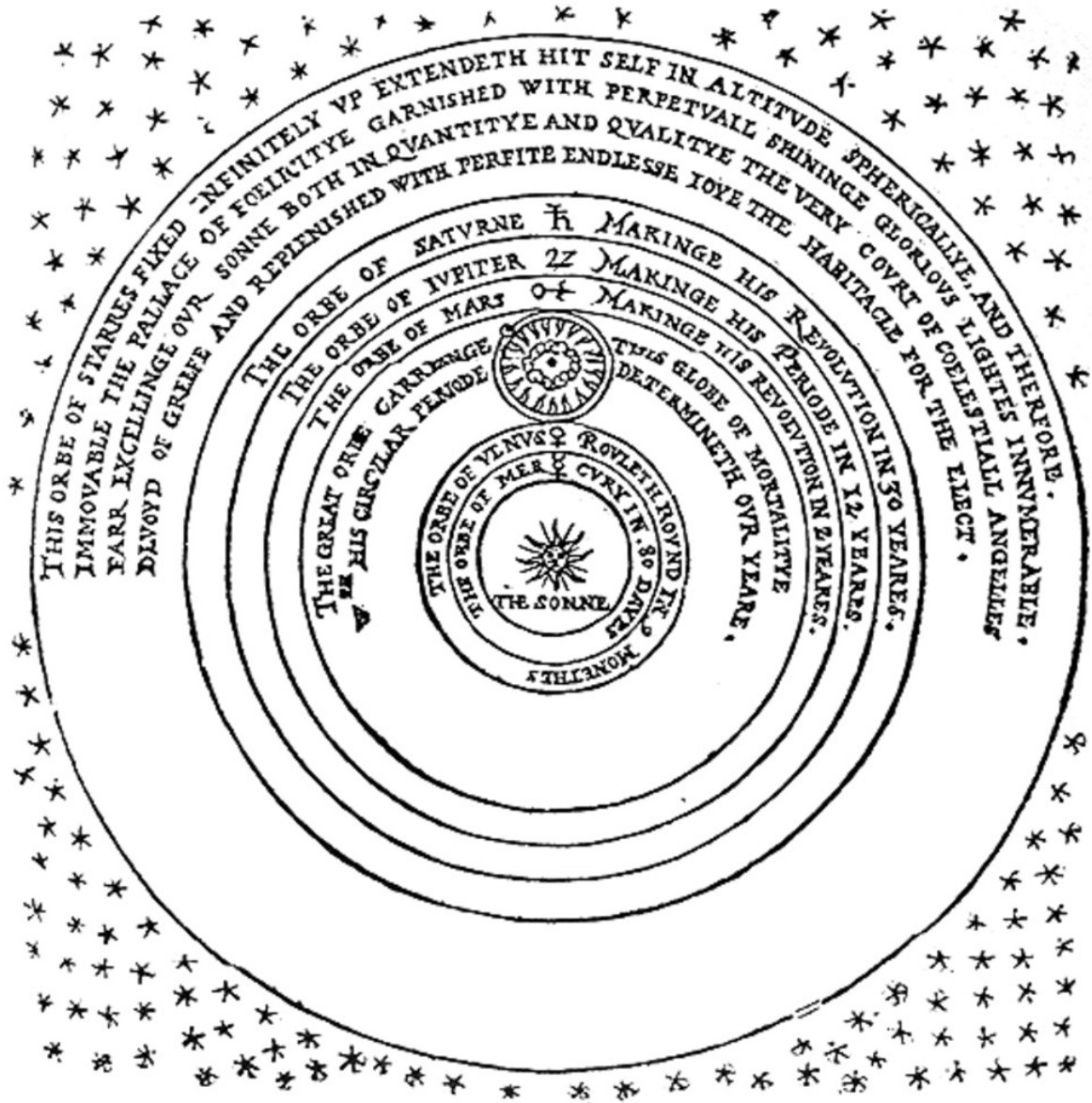

FIGURE 2: The Copernican System as portrayed in Thomas Digges' *A Perfit Description of the Coelestiall Orbes* (1576) with no firmament and the stars extending indefinitely into space. The idea that the distances between these stars is large appears, for example, in Giordano Bruno's 1584 *La Cena de le Ceneri*.

> star of fifth or sixth magnitude seems to equal the Dog Star,
> largest of all the fixed stars [p. 46].

However, there is plenty of evidence that within a few years of the publication of the *Starry Messenger* Galileo came to view stars as being defined, spherical, consistently measurable celestial bodies -- bodies that he believed were suns at varying distances in space.  Three examples show that Galileo held this view of the stars over a period of at least 15 years.

   The first example is that in 1617 he observed and split the double star Mizar (Ondra 2004, Siebert 2005).  His observing notes contain the following measurements -- separation: 0°, 0', 15"; larger star radius: 0°, 0', 3"; smaller star radius -- 2"; gap between them -- 10".  He also notes that the radius of the sun contains 300 radii of the larger star, so therefore the distance to the star contains 300 distances to sun, if the star is the size of the sun (*Opere*, III, p. 877).[3]  Here we see that Galileo has concluded that stars do have a circular periphery.  He can measure their radii (or what he thinks is their radii).

---

3 The text of Galileo's notes:

> *Inter mediam caudae Elicis et sibi proximam pond nunc gr. 0.0'.15".*
> *Semidiameter stellae maioris gr. 0.0.3", minoris vero 2", et intercapedo 10"....*
> *Semidiameter ☉ continet semidiametros stellae maioris 300. Distantia ergo stellae continet distantias ☉ 300 (si stella ponatur tam magna ut ☉)....*

  Elicis is Helice -- the Great Bear.

He is doing calculations based on the assumption that they are suns.

The second example is that in 1624 Galileo wrote a lengthy letter to Francesco Ingoli in which he makes arguments that stars are defined, spherical, consistently measurable celestial bodies that are suns scattered throughout space.  He does this in the course of refuting opponents of Copernicus:

> These opponents of Copernicus make certain calculations based on the premise that, although the earth's motion in its annual orbit produces some curious and extremely large changes in the case of the planets, it does not cause any similar effects in the case of the fixed stars; they calculate that the stellar sphere would have to be so far away that a fixed star would itself have to be many times larger than the whole annual orbit in order for it to be visible to us with the magnitude with which it appears to us; this in turn would mean a size many thousands of times bigger than the sun itself, which they regard as the greatest absurdity. However, my calculations show me that this business proceeds very differently, namely that taking an average fixed star to be as large as the sun and no larger is sufficient to solve all difficulties which, through their own errors, they have attributed to Copernicus.  Their errors were to consider the apparent magnitudes of the stars (fixed as well as wandering) much greater than what they are....  I say that if you measure Jupiter's diameter exactly, it barely comes to 40 seconds, so that the sun's diameter becomes 50 times greater; but Jupiter's diameter is no less than ten times larger than that of an average fixed star (as a good telescope will show us), so that the sun's diameter is five hundred times that of an average fixed star; from this it immediately follows that the distance to the stellar region is five hundred times greater than that between us and the sun. Now, what would you expect if the earth is displaced from

> the center of the stellar sphere by one or two parts out of five
> hundred, in regard to whether stars appear smaller at the horizon
> than at the meridian?  Who will be so simpleminded as to believe
> that ordinary astronomers can detect such a small increase or
> decrease in the diameter of a star, when we can grasp with our
> hands that in similar observations they have been deceived so
> seriously, as I mentioned above?  Thus, the objections of
> opponents are removed, as you see, simply by taking fixed stars,
> for example those of the third magnitude, to be equal in size to
> the sun.  However, since with the telescope we see countless
> others that are much smaller than even those of the sixth
> magnitude, since we can reasonably believe there are many others
> which are not observable with the telescopes built so far, and
> since there is no difficulty in believing that they are equal in
> size and occasionally larger than the sun, at what remote
> distance do you feel we can say without difficulty that they must
> be located?  Fixed stars, Mr. Ingoli, give off their own light,
> as I have proved elsewhere; so they lack nothing to be entitled
> to be called and considered suns [Reply to Ingoli, pp.
> 167-168]....

Later in the letter Galileo again states his ideas on measuring the sizes of fixed stars:

> ...it is already clear that you and Tycho make the stellar
> sphere sixty times greater than required to save Copernicus's
> position, and that you do so somewhat arbitrarily and not
> because you have examined carefully the apparent magnitude of
> fixed stars.  This is a pruning or subtraction of no small
> moment, I mean to cut by ninety-eight percent the size you
> condemned.  Nor can it be true, if you do not mind, that I
> ever said a fixed star subtends 2 minutes, as you claim I have;
> for it was many years ago that I learned by sensory experience
> that no fixed star subtends even 5 seconds, many not even 4, and

> innumerable others not even 2 [Reply to Ingoli pp. 173-174].

But he is still not finished, again hammering home the idea that the stars are scattered through space:

> I do not think the fixed stars are are all placed on a spherical surface, so as to be equidistant from a particular point, such as the center of a sphere; indeed only God knows whether for any group larger than three there is a single point from which they are equidistant [Reply to Ingoli, p. 176]

He also re-emphasizes that the stars are globes:

> ...why do you not place us in the firmament, where there are centers by the thousands, since every star is a perfect globe and every globe has its own center? [Reply to Ingoli, p. 180]

The third example is Galileo's 1632 *Dialogue*. As noted earlier, in the *Dialogue* he again states the ideas that he had expressed in 1617 and 1624 -- that stars are suns (p. 327) distributed through space (p. 382). And again he gives measurements of stellar diameters and uses those measurements to calculate distances to stars:

> ...I shall show that by assuming that a star of the sixth magnitude may be no larger than the sun, one may deduce by means of correct demonstrations that the distance of the fixed stars from us is sufficiently great to make quite imperceptible in them the annual movement of the earth which in turn causes such large and observable variations in the planets....

> ...I assume with the same concurrence and in accordance with the
> truth that the apparent diameter of the sun at its average
> distance is about one-half a degree, or 30 minutes; this is 1,800
> seconds, or 108,000 third-order divisions.  And since the
> apparent diameter of a fixed star of the first magnitude is no
> more than 5 seconds, or 300 thirds, and the diameter of one of
> the sixth magnitude measures 50 thirds (and here is greatest
> error of Copernicus's adversaries), then the diameter of the sun
> contains the diameter of a fixed star of the sixth magnitude
> 2,160 times.  Therefore if one assumes that a fixed star of the
> sixth magnitude is really equal to the sun and not larger, this
> amounts to saying that if the sun moved away until its diameter
> looked to be 1/2160th of what it now appears to be, its distance
> would have to be 2,160 times what it is in fact now.  This is the
> same as to say that the distance of a fixed star of sixth
> magnitude is 2,160 radii of the earth's orbit [*Dialogue* p.
> 359-360].

Looking at these three examples it is clear that the idea that the stars are suns was a prominent and recurring feature in Galileo's thinking as regards the stars -- or put another way, the Copernican Principle idea that the sun is but one of many suns was a prominent and recurring feature in Galileo's thinking as regards the stars.  It is also clear that Galileo believed that he could reliably measure the angular diameters of stars, that those diameters represented the actual sizes of the stars, and that therefore his measurements in conjunction with the assumption that the stars were suns could be used to determine stellar distances, which he consistently reported as ranging from several hundred AU for brighter stars to over two thousand AU for fainter stars.[4]

Today we know that these ideas were wrong. All stars are not the same size as the sun. The star images Galileo measured of course did not reflect the true sizes of the stars. The distances Galileo determined were orders of magnitude too small. However, while Galileo's ideas are wrong, they are not illogical.

An anti-Copernican of Galileo's time could also claim that Galileo's ideas about the stars were wrong. An anti-Copernican could claim that the stars were, in fact, all placed on a spherical surface which rotated about Earth daily. Prior to Galileo, Tycho Brahe had attempted, via observations of Mars, to settle the question of whether the heliocentric Copernican or the geocentric Ptolemaic model of the universe was correct; unable to do so, he ended up developing a geocentric model of his own in which the planets circle the sun while the sun circles the Earth -- a model which has the advantages of Copernicus's ideas without the problems posed by a moving Earth (Gingerich and Voelkel 1998; Figure 3). Galileo's observations of the phases of Venus and the moons of Jupiter undercut the Ptolemaic geocentric model but not the Tychonic one, while his theory of the tides and other efforts to support the Copernican theory were flawed (Graney 2008).  The Tychonic and Copernican models were observationally identical insofar as the planets were concerned; only observations of the stars, such as the successful detection of annual stellar parallax, could yield direct support for the Copernican model.  Well after Galileo's death, the issue of whether the heliocentric Copernican model or the geocentric Tychonic one was correct remained unsettled (Figure 4) -- as late as the 1670's Robert Hooke argued that the issue could not be settled absent a detection of parallax (Hooke 1674).  Direct observational evidence for Earth's motion, if not for the Copernican Principle in general, did not arrive until nearly a century after the publication of the *Dialogue*, when James Bradley detected the aberration of starlight due to the relative motion of the Earth and stars in 1728.

---

4 The reader skeptical that Galileo could realistically measure such small sizes should keep in mind the 1617, 1624, and 1632 statements about star sizes already mentioned, and a measurement of Sirius as being 5" in diameter (*Opere* III, Pt 2, p. 878). Galileo believes he can distinguish 3" from 2", and 5" from 4" from 2". Drake and Kowal (1980) note Galileo recorded a change in the apparent size of Jupiter from 41.5" to 39.25" and that his measurements are consistent with modern calculations. Standish and Nobili (1997) and Graney (2007) note the accuracy of Galileo's measurements and sketches. Graney (2007) estimates that accuracy at 2".

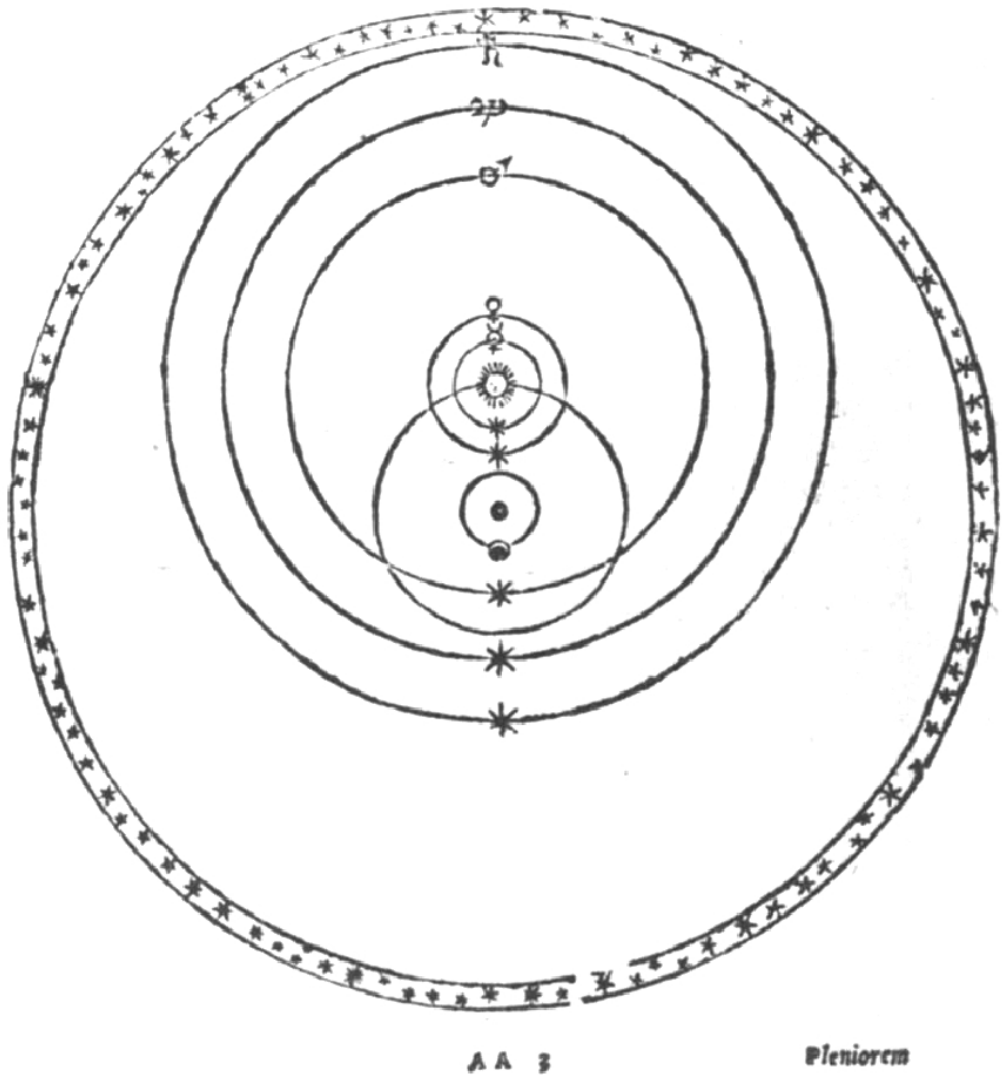

**FIGURE 3:** The Tychonic System from Tycho's *De Mundi Aetherei Recentioribus Phaenomenis* (1588). The stars are on a firmament.

3. A COPERNICAN PRINCIPLE THOUGHT EXPERIMENT

A simple thought experiment using Galileo's methods will illustrate that the Copernican Principle had observational consequences aside from parallax (or aberration).  These consequences could have been tested by Galileo.  What's more, given Galileo's logical but erroneous ideas about the stars, the results of such testing would have appeared to confirm the Copernican Principle.

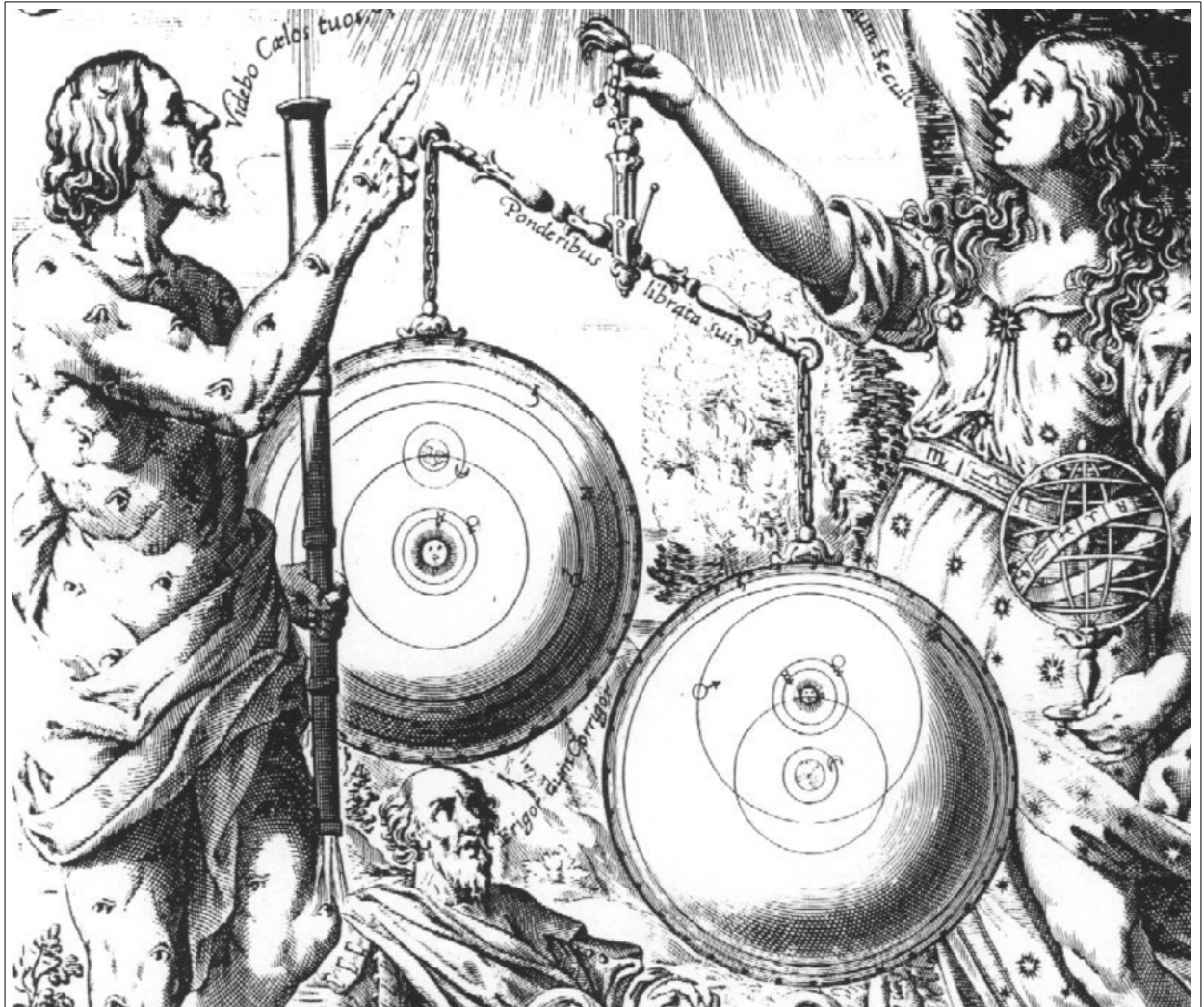

**FIGURE 4: Illustration from Giovanni Battista Riccioli's *Almagestum Novum* (1651) showing how the accepted theory of the universe remained unsettled after Galileo. The Copernican heliocentric theory (left side of the balance) is being weighed against a geocentric theory similar to that of Tycho (right side). The scales are tipped in favor of the geocentric theory, indicating that it is the better of the two in Riccioli's view. Both theories are shown with firmaments of stars.**

Let us begin with the Copernican Principle as expressed by Galileo -- the stars are so many suns, scattered over varying distances. The number of stars $N^*$ located within a radius $r$ of Earth is then

(1) $\qquad N^*(r) = 4/3\ \pi\ \rho^*\ r^3$

where $\rho^*$ is the average number density of stars in space.

Next we move to Galileo's determination of stellar distances. Following Galileo's previously mentioned work of 1617, 1624, and 1632, we can write that Galileo figured the distance $L$ to a star whose angular diameter he measured to be $\alpha$ as $L = \alpha_\odot/\alpha$, where $\alpha_\odot$ is the angular diameter of the sun and $L$ is measured in AU. Combining this with equation 1 gives us an equation for the number of stars with apparent angular diameter of $\alpha$ or greater:

(2) $\qquad N^*(\alpha) = 4/3\ \pi\ \rho^*\ (\alpha_\odot/\alpha)^3$

So far our thought experiment has consisted of the Copernican Principle and basic geometry.

For the final step in our thought experiment, we proceed to Galileo's ideas about the relationship between a star's telescopically measured angular diameter $\alpha$, and its magnitude as determined with the naked eye $M$. In the *Dialogue* Galileo states that stars of $M = 1$ have $\alpha = 5''$ and stars of $M = 6$ have $\alpha = 5/6''$ (p. 359), suggesting $M$ and $\alpha$ are related by

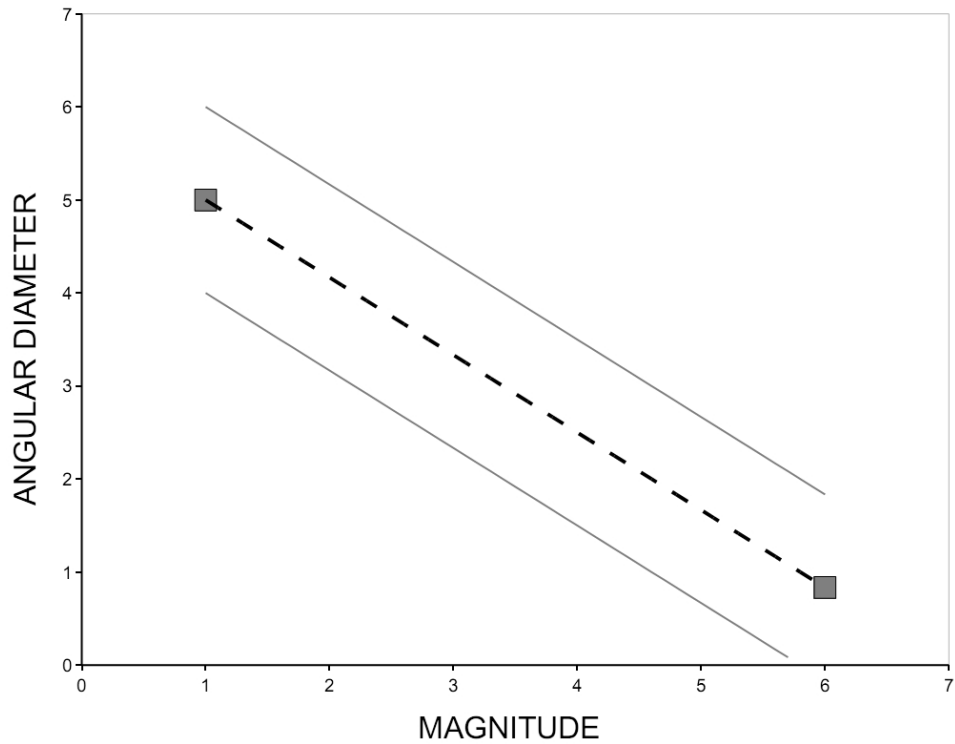

**FIGURE 5:** Plot showing Galileo's values of magnitude *M* and angular size $\alpha$ from the *Dialogue* (marked points). Dotted line is linear relationship between *M* and $\alpha$. Thin solid lines represent Galileo's measurement error.

(3)  $\alpha = (5/6)(7 - M)$

(Figure 5)[5]; this substituted into equation 2 yields the following for *N\** as a function of *M*:

(4)  $N^*(M) = 288/125\ \pi\ \rho^*\ (\alpha_\odot/(7-M))^3$

---

5 Graney (2007) concludes that the linear relationship between $\alpha$ and *M* implied in the *Dialogue* is consistent with what would be seen through the telescope, which could suggest that Galileo's 5/6" diameter for 6$^{th}$ magnitude stars is a projection from a trend.

(Figure 6).  Here $N*$ is the number of naked eye stars visible from Earth of magnitude $M$ or brighter.  Our thought experiment has led us to something that Galileo would think he could use as a test of his ideas -- if a count of naked eye stars yields data consistent with equation 4, then the Copernican Principle gets some observational support.

4.  COPERNICAN PRINCIPLE PREDICTIONS "CONFIRMED"

Data from the *Bright Star Catalog* on the numbers of visible stars brighter than a given magnitude are given in Figure 7 (Hoffleit 1991).  Proceeding from Galileo's assessment that $6^{th}$ magnitude stars lie at a distance of *2160 AU*, stars $6^{th}$ magnitude and brighter would be contained within a spherical volume of *42.2 x 10$^9$ AU$^3$*, and $\rho* = 8404/42.2 \times 10^9 \text{ AU}^3 = 1.991 \times 10^{-7}$ *star/AU$^3$*.  Using this value with equation 4 to plot $N*$ vs. $M$ yields a result consistent with the *Bright Star Catalog* data (Figure 8).

Of course Galileo did not have access to the *Bright Star Catalog*, but obtaining a good estimate of such data is simply a matter of doing a count of naked eye stars by magnitude in various sections of the sky, and then extrapolating to numbers for the sky as a whole.  Some such data was already available -- Ptolemy's *Almagest* contained a catalog of over a thousand stars with brightness rated from first magnitude[6] on down to very faint stars (although the catalog contained comparatively few faint stars). This data could serve as a starting point for a more thorough star count by Galileo; it shows that such counts could be done; and it even hints at support for the Copernican Principle (Figure 9). Counting stars with the naked eye would be a tedious and time-consuming task, no doubt, but not requiring skill or stamina beyond that shown by Galileo, who, for example, recorded precise sketches of the Jovian system on a daily and even hourly basis for more than two months (*Opere* V, p. 241) and figured out how to achieve 2" accuracy in measurements (Graney 2007) using simply a grid of paper as a measuring device (Drake and Kowal 1980).

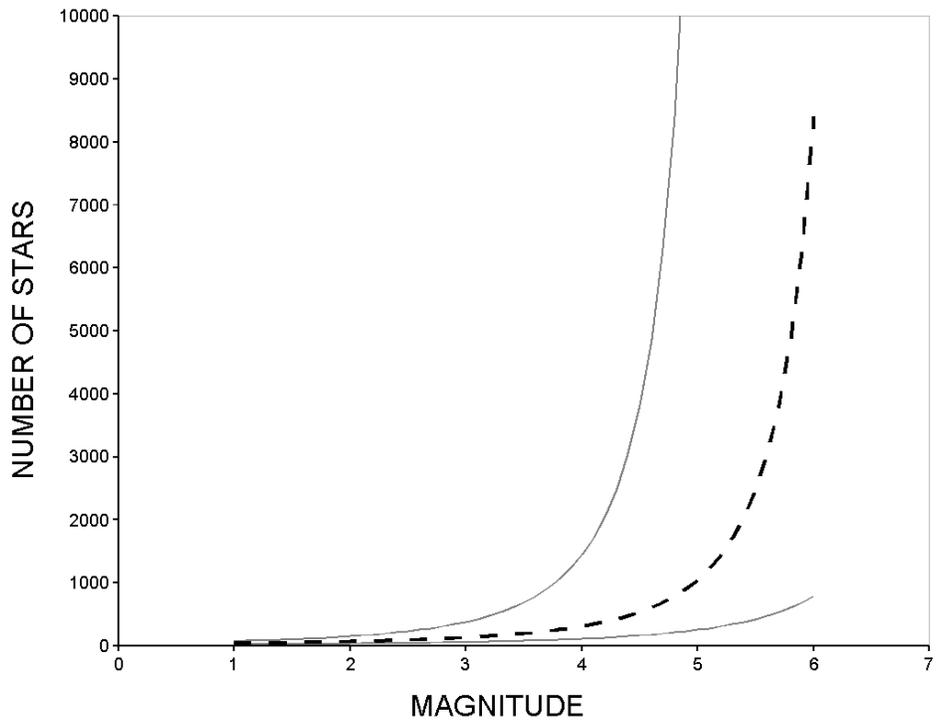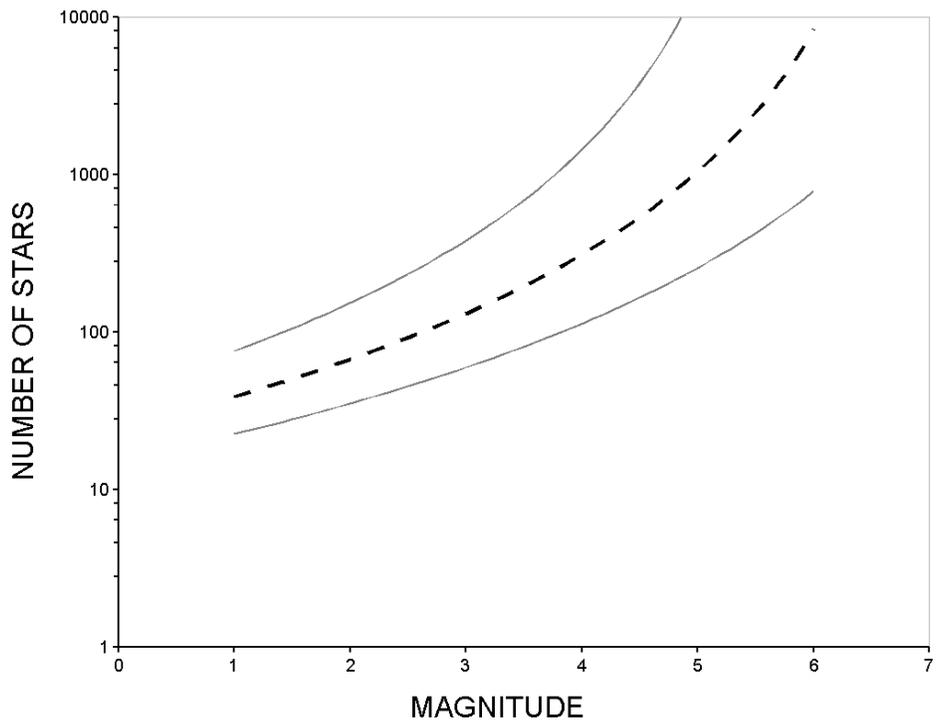

FIGURE 6: Plot of equation 4 -- linear (top) and log (bottom) scale. The thin solid lines show allowance for measurement error.

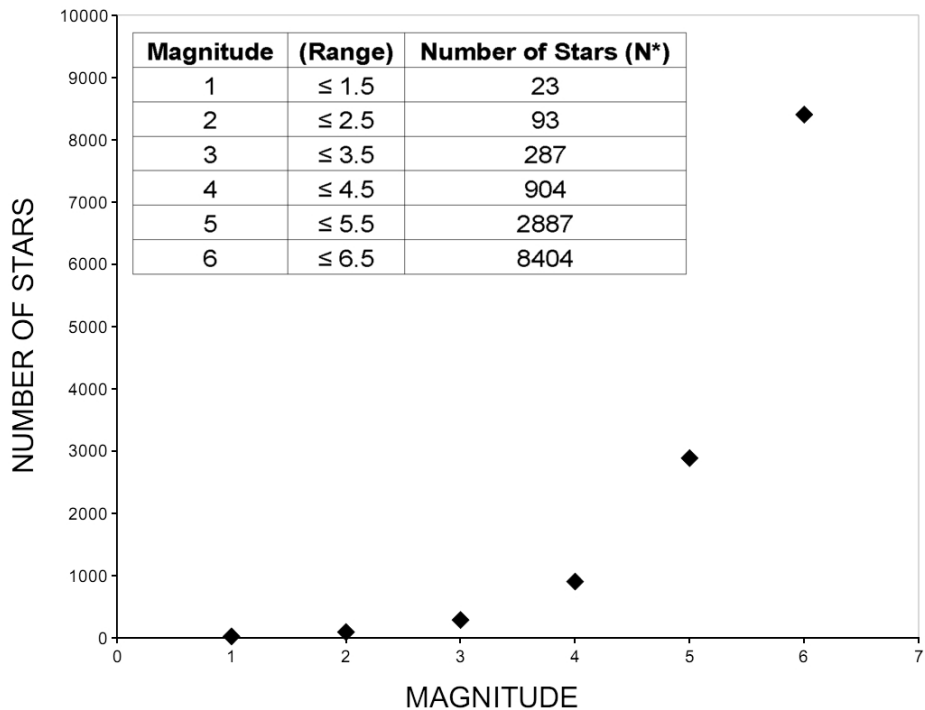

FIGURE 7: Table and plot of numbers of stars and magnitudes from *Bright Star Catalog* (Hoffleit 1991).

---

6 The reader should note that these included stars that would be rated as having zero or negative magnitude by the modern system. For example, Sirius, Arcturus, and Spica are all first magnitude according to the *Almagest* catalog.

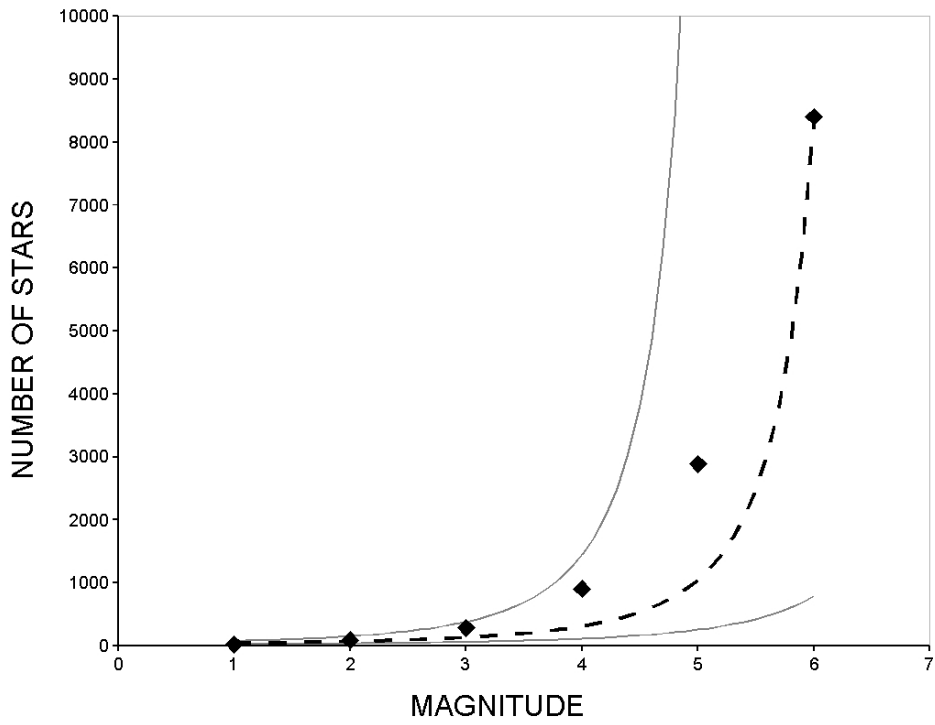
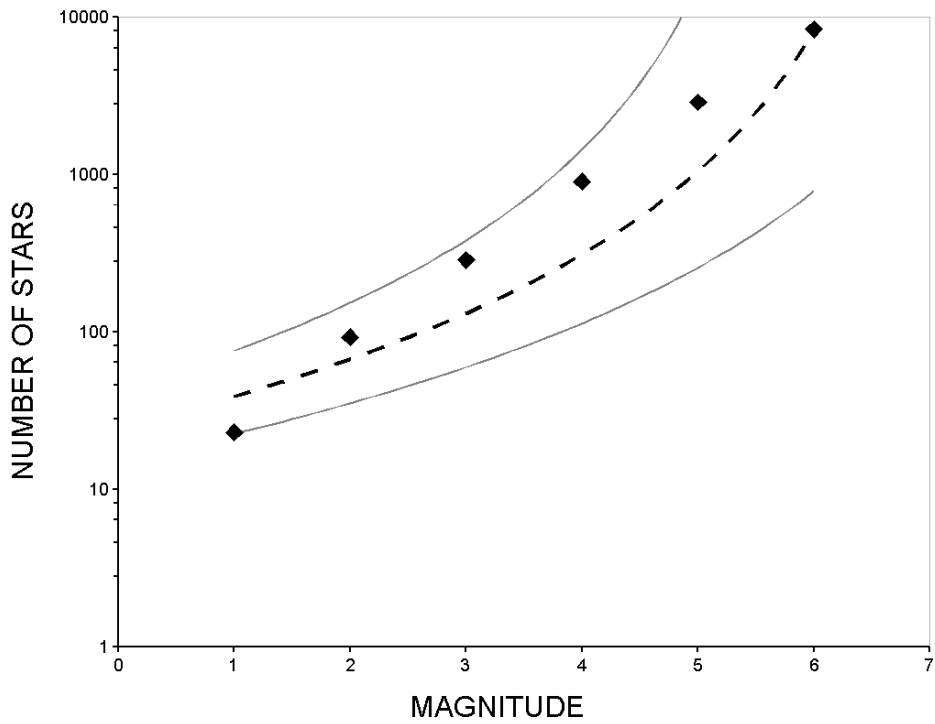

FIGURE 8: Data from *Bright Star Catalog* plotted with equation 4 -- linear (top) and log (bottom) scale.

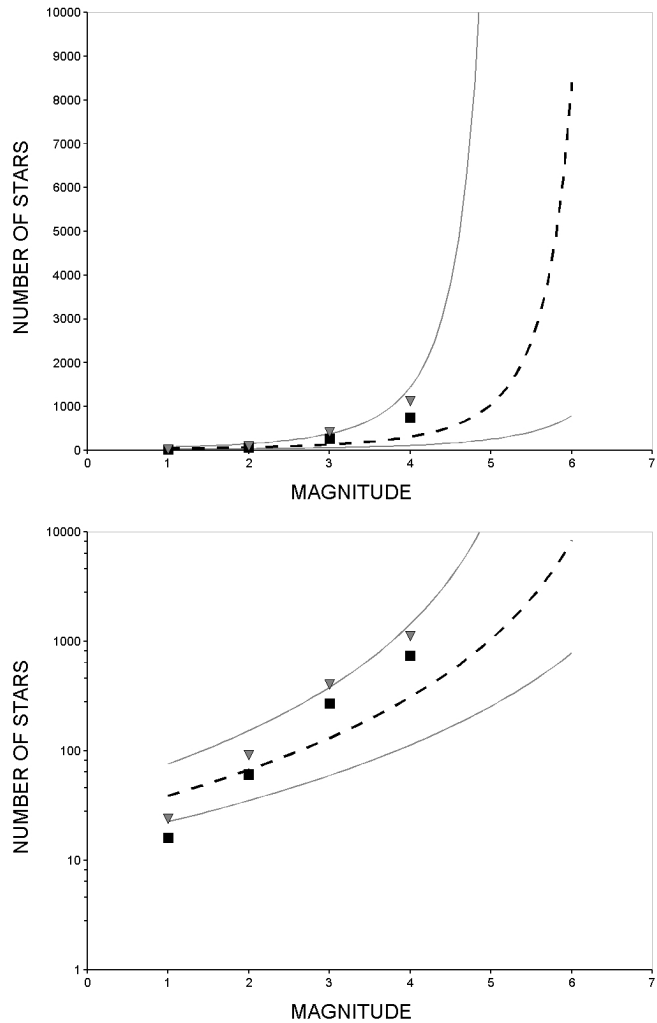

FIGURE 9: Data from the *Almagest* catalog (Jaschek 1987) plotted with equation 4 -- linear (top) and log (bottom) scale. The *Almagest* catalog shows increasing numbers of stars with magnitude up to 4$^{th}$ magnitude after which the numbers decrease. Since a good look at the night sky indicates that there are more fainter stars than brighter ones, it appears the A*lmagest* did not intend to thoroughly catalog faint stars; therefore for this plot the *Almagest* data was cut off at 4$^{th}$ magnitude. Stars were grouped as in Figure 7 (magnitude 1 includes all stars of magnitude 1.5 or brighter, etc.). Square points are directly from the *Almagest*. Triangular points are *Almagest* data extrapolated to the entire sky (multiplying by 1.5 so as to approximately match the *Bright Star Catalog* data for 1$^{st}$ and 2$^{nd}$ magnitude stars).

In short, the naked-eye appearance of the stars in the night sky speaks to Galileo, with his erroneous but logical ideas about stars, in support of the Copernican Principle. In light of this thought experiment, any anti-Copernican who would claim that the stars are all placed on a spherical surface which rotates about Earth daily faces the task of having to explain why it happens to be that *N\** increases with magnitude in a way so consistent with the Copernican Principle. After all, if the stars are simply bodies located on a firmament, any relationship between N* and M is possible. There could be equal numbers of each magnitude of star. There could be more bright stars than faint stars. So why is it that the appearance of the naked eye stars matches that which would be expected in a Copernican Principle universe of suns extending indefinitely out into space (Figure 10)?

Recall from earlier in this paper that in the "Reply to Ingoli" Galileo suggests that the stars extend indefinitely out into space, saying that the telescope reveals countless stars much "smaller" (fainter) than sixth magnitude, and that it is reasonable to believe there are many others which are not observable with the telescopes available (p. 167-168). Thomas Digges had assumed this sort of universe of stars (Figure 2) and argued that such an infinite universe had to be immobile. Our thought experiment backs up Digges' (and Galileo's) ideas in this regard, arguing for the immobility of the stars. However, we have to acknowledge that it does not quite prove the motion of the Earth about the sun -- a stubborn anti-Copernican could argue that immobile stars only requires the Earth to rotate; the sun and planets could revolve about the Earth as in the Tychonic system. Nonetheless our thought experiment would be a powerful argument for Copernican ideas in general -- with our thought experiment in hand, Galileo could simply look up at the night sky and say "this *looks* like a Copernican universe"!

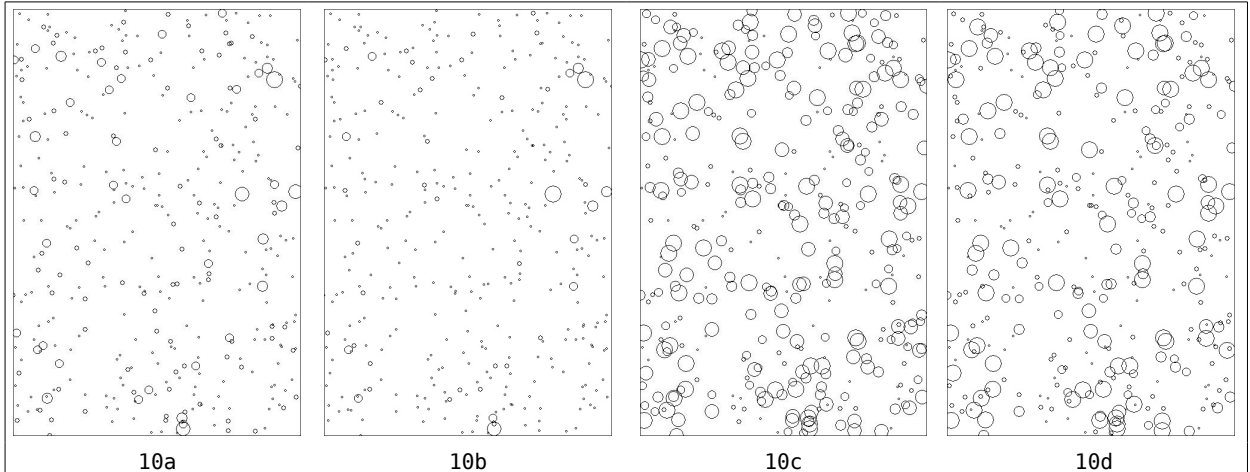

FIGURE 10: Simulated field of stars of magnitudes 1 through 6 (larger circles representing brighter stars). 10a -- numbers of each magnitude in proportions found in *Bright Star Catalog* (i.e. real sky). 10b -- numbers calculated via equation 4. 10c -- equal numbers of each magnitude. 10d -- numbers of each magnitude selected at random. If stars are not suns scattered through space then there is no reason for the real sky to look like 10a and 10b. For example, if the stars are simply bodies distributed along a spherical shell centered on Earth as in geocentric theories then there is no reason why their numbers by magnitude might not be equal or even random (with no particular relationship between magnitude and number of stars).

5.  CONCLUSION

Our thought experiment illustrates a possible method for Galileo to obtain experimental evidence to support his Copernican ideas; hopefully it may prompt further research into whether Galileo actually investigated this line of thought.  Galileo remained a staunch Copernican even though the Tychonic system provided a viable explanation of the available observations without the problems created by a moving Earth (such as the expectation of seeing parallax).  Finding an observational test in support of the Copernican view could help further illuminate why Galileo supported Copernicus so strongly (not even giving the Tychonic system the nod that Hooke gave it) -- for by supporting Copernicus so strongly Galileo certainly helped to move astronomy forward.

   As outlined in this paper, the night sky can confirm the Copernican Principle -- at least to an observer such as Galileo who has established a (flawed) $17^{th}$ century understanding of the apparent relationship between stars' magnitudes and their angular sizes seen through the telescope.  Viewed with this flawed understanding, the manner in which the number of visible stars increases with magnitude supports the Copernican Principle that the stars are suns scattered throughout space.  The ideas of later $17^{th}$ century astronomers such as Hooke suggest that this idea did not occur to them, as the Tychonic model remained viable well after Galileo's death.  Whether the idea ever occurred to Galileo, and if so, what impact it had on his views, could be a subject for further research by those with good access to original sources.

ACKNOWLEDGEMENTS

I would like to thank an anonymous referee for a tough-minded but helpful critique of this paper.